\documentclass[preprint,epsf,preprintnumbers,nofootinbib]{revtex4-1}
\usepackage{graphicx}
\usepackage{dcolumn}
\usepackage{bm}
\usepackage{amsmath}
\usepackage{amsfonts} 
\usepackage{latexsym}
\usepackage{bbm}
\usepackage{color}
\newcommand{\sumint}{\mbox{$\sum$}\kern-2.7ex\int}

\begin{document}

\preprint{CTPU-PTC-18-24}

\title{Revisiting electroweak phase transition in the standard model with a real singlet scalar}

\author{Cheng-Wei Chiang$^{1,2,3}$}%
\email{chengwei@phys.ntu.edu.tw}
\author{Yen-Ting Li$^{1}$}%
\email{R04222022@ntu.edu.tw}
\author{Eibun Senaha$^{4}$}%
\email{senaha@ibs.re.kr}

\affiliation{$^1$Department of Physics, National Taiwan University, Taipei 10617, Taiwan}
\affiliation{$^2$Institute of Physics, Academia Sinica, Taipei 11529, Taiwan}
\affiliation{$^3$Kavli IPMU, University of Tokyo, Kashiwa, 277-8583, Japan}
\affiliation{$^4$Center for Theoretical Physics of the Universe, Institute for Basic Science (IBS), Daejeon 34126, Korea}

\bigskip

\date{\today}

\begin{abstract}
We revisit the electroweak phase transition in the standard model with a real scalar,
utilizing several calculation methods to investigate scheme dependences.
We quantify the numerical impacts of Nambu-Goldstone resummation, required in one of the schemes, 
on the strength of the first-order electroweak phase transition.
We also employ a gauge-independent scheme to make a comparison with the standard gauge-dependent results.
It is found that the effect of the Nambu-Goldstone resummation is typically $\sim1\%$.
Our analysis shows that both gauge-dependent and -independent methods
give qualitatively the same result within theoretical uncertainties.
In either methods, the scale uncertainties in the ratio of critical temperature and the corresponding Higgs vacuum expectation value are more than 10\%, which signifies the importance of higher-order corrections. 
\end{abstract}


\maketitle

\section{Introduction}

Cosmic baryon asymmetry~\cite{Patrignani:2016xqp} is one of the longstanding problems in particle physics and cosmology.
Though the standard model (SM) can satisfy the so-called the Sakharov criteria~\cite{Sakharov:1967dj} in principle, 
the discovered Higgs boson with a mass of 125 GeV~\cite{125h} is incompatible with successful electroweak baryogenesis (EWBG)~\cite{ewbg} 
since the electroweak phase transition (EWPT) is a smooth crossover~\cite{lattice} rather than 
first order with expanding bubbles.  
It is known that this drawback can be easily circumvented by augmenting the minimal Higgs sector.
The simplest extension is to add an SU(2)$_L$ singlet scalar, 
which provides not only a strong first-order EWPT but also a dark matter candidate 
if a $Z_2$ symmetry is imposed ~\cite{EWPT_rSM,Curtin:2014jma,Curtin:2016urg,EWPT_DM_rSM,HT_xSM,DM_rSM,DM_rSM_recent}.

A thorny problem in investigating EWPT using a perturbative effective potential is the dependence on a
gauge fixing parameter $\xi$~\cite{Jackiw:1974cv,Dolan:1974gu} (for recent studies, see, {\it e.g.}, Refs.~\cite{Patel:2011th,Garny:2012cg}).
For instance, the Higgs vacuum expectation value (VEV) obtained by the
effective potential can change with a varying $\xi$.
Such an unwanted $\xi$ dependence eventually contaminate a baryon-number preserving criterion: 
$v_C/T_C\gtrsim 1$, where $T_C$ denotes the critical temperature associated with the phase transition and $v_C$ is the doublet Higgs VEV at $T_C$.
As a result, any phenomenological consequences derived from this criteria suffer from 
the $\xi$ dependence and are therefore unreliable unless the dependence can be kept under control.

Common lore is that if the EWPT is driven by scalar thermal loops or a tree-potential barrier, 
the $\xi$ dependence is expected to be small.
As found in the Abelian-Higgs model with an additional scalar~\cite{Wainwright:2012zn}, 
however, such an expectation is not always correct. 
It is concluded that the $\xi$ dependence can be pronounced even when the tree-potential barrier exists.
Nevertheless, this point is often overlooked in previous studies on the EWPT in the SM with a real singlet scalar.

Another issue is the occurrence of IR divergences in the effective potential in the $R_\xi$ gauge with $\xi=0$. For example, if the Higgs boson mass is renormalized using the one-loop effective potential in such a way that the loop corrections do not modify the tree-level mass relations, 
\footnote{This is called ``on-shell" renormalization in Ref.~\cite{Curtin:2014jma}. Since it is not 
the genuine on-shell renormalization, we refer to it as ``on-shell-like" renormalization in the current paper. }
the second derivative of the one-loop effective potential is ill-defined due to the IR divergences
coming from the Nambu-Goldstone (NG) boson loops.
One of the prescriptions for the problem is to resum higher-order corrections to the NG masses~\cite{Elias-Miro:2014pca,Martin:2014bca}.
One can show that the NG contributions have little effect on the Higgs mass once they are resummed.
Nonetheless, it would be desirable to quantify their numerical impact on $v_C/T_C$ explicitly.

In this paper, we revisit EWPT in the SM with a singlet scalar, focusing on the aforementioned two issues as well as the scheme dependence. 
We first clarify the numerical importance of the thermal gauge boson loops on $v_C/T_C$ by subtracting them off from the finite-temperature effective potential in the Landau gauge $\xi=0$. 
Even though this simple method cannot precisely quantify the $\xi$ dependence, it tells how important 
the thermal gauge loops can be in order to achieve a strong first-order EWPT, 
especially $v_C/T_C\simeq 1$.
We regard this as a simple criterion whether a further investigation of the $\xi$ dependence
is needed or not.

In addition to the numerical studies of $v_C/T_C$ in the on-shell (OS)-like scheme with the 
NG resummation, we also evaluate $v_C/T_C$ utilizing the following three methods commonly adopted
in the literature for comparison:
(1) the $\overline{\text{MS}}$ scheme, (2) the high-temperature (HT) potential defined as the
tree-level potential plus thermal masses and (3) the Patel-Ramsey-Musolf (PRM) scheme~\cite{Patel:2011th}.
In the first method, the tree-level NG boson masses are not zero
in the one-loop corrected vacuum so that the NG resummation mentioned above is not required.
The second method is manifestly gauge invariant since the thermal masses do not have the $\xi$ dependence. 
In the last one, the Nielsen-Fukuda-Kugo (NFK) identity~\cite{Nielsen:1975fs,Fukuda:1975di} is used to obtain the gauge-invariant $T_C$, and $v_C$ is determined by use of the HT potential. 
We confine ourselves to $\mathcal{O}(\hbar)$ calculations
in which the thermal resummation is not performed. 
Going beyond this order requires two-loop contributions as well, which is out of the scope of current investigation. 
Application of the $\mathcal{O}(\hbar)$ PRM scheme to the SM with a complex scalar can be found in Ref.~\cite{Chiang:2017nmu}. However, devoted numerical comparisons between this scheme
and the standard gauge-dependent ones are not performed. 
One of the goals of this study is to complement this part.

The paper is organized as follows.
In Sec.~\ref{sec:model}, we introduce the model and define our notation.
Renormalization schemes are given to fix the input parameters. 
In Sec.~\ref{sec:ewpt}, we outline the EWPT in the model.
Sec.~\ref{sec:results} shows the results of our numerical analyses. 
The conclusion and discussions are given in Sec.~\ref{sec:conclusion}.

\section{Model}\label{sec:model}

We consider a model in which an $\text{SU}(2)_L$ singlet real scalar $S$ is added to the SM.
The $S$ boson can be a dark matter candidate if a $Z_2$ symmetry is imposed~\cite{DM_rSM}. 
The tree-level Higgs potential with the $Z_2$ symmetry is then cast into the form:
\begin{align}
V_0(H, S) = -\mu_H^2H^\dagger H+\lambda_H(H^\dagger H)^2-\frac{\mu_S^2}{2}S^2
	+\frac{\lambda_S}{4}S^4+\frac{\lambda_{HS}}{2}H^\dagger HS^2.
\end{align}
The doublet Higgs field is parametrized as
\begin{align}
H(x)=\left(
\begin{array}{c}
G^+(x) \\
\frac{1}{\sqrt{2}}\big[v+h(x)+iG^0(x)\big]
\end{array}
\right),
\end{align}
where $v\simeq 246$ GeV denotes the VEV, $h$ the 125-GeV Higgs boson, and $G^{0,\pm}(x)$ the NG bosons.

The tadpole conditions at tree level are
\begin{align}
\begin{split}
T_h &\equiv \left\langle\frac{\partial V_0}{\partial h}\right\rangle = 
v\Big[-\mu_H^2+\lambda_H v^2+\frac{\lambda_{HS}}{2}v_S^2\Big]=0,\\ 
T_S &\equiv \left\langle\frac{\partial V_0}{\partial S}\right\rangle = 
v_S\left[-\mu_S^2+\lambda_S v_S^2+\frac{\lambda_{HS}}{2}v^2\right]=0,
\end{split}
\end{align}
where the symbol $\langle\cdots\rangle$ means that the quantity sandwiched by the angled brackets 
is evaluated in the vacuum, and $v_S=\langle S\rangle$.
The $Z_2$-invariant vacuum corresponds to the solution: $\mu_H^2=\lambda_Hv^2$ and $v_S=0$,
from which the scalar boson masses are given by
\begin{align}
\begin{split}
m_h^2 &=-\mu_H^2+3\lambda_Hv^2=2\lambda_H v^2,\\
m_S^2 &= -\mu_S^2+\frac{\lambda_{HS}}{2}v^2.
\end{split}
\end{align}
Denoting the background fields of $H$ and $S$ as $\varphi / \sqrt2$ and $\varphi_S$, respectively, 
the tree-level effective potential takes the form
\begin{align}
V_0(\varphi, \varphi_S) &= -\frac{\mu_H^2}{2}\varphi^2+\frac{\lambda_H}{4}\varphi^4 
	+\frac{\lambda_{HS}}{4}\varphi^2\varphi_S^2-\frac{\mu_S^2}{2}\varphi_S^2
	+\frac{\lambda_S}{4}\varphi_S^4.
\end{align}
To avoid an unbounded-from-below potential,
one has to have $\lambda_H>0$ and $\lambda_S>0$, and additionally $-2\sqrt{\lambda_H\lambda_S}<\lambda_{HS}$ if $\lambda_{HS}<0$. As far as the strong first-order EWPT is concerned, 
$\lambda_{HS}>0$ is necessary so that the last condition is irrelevant in our study.

For $\mu_S^2>0$, a local minimum can appear 
in the singlet scalar direction (denoted as $v_S^{\text{sym}}$) before electroweak symmetry breaking (EWSB).
For the EW vacuum to be the global minimum after the EWSB, one must have
\begin{align}
V_0(v, 0) < V_0(0, v_S^{\text{sym}})
\quad \Longrightarrow\quad \lambda_S
>\lambda_H\frac{\mu_S^4}{\mu_H^4}
=\frac{2}{m_H^2v^2}\left(m_S^2-\frac{\lambda_{HS}}{2}v^2\right)^2\equiv \lambda_S^{\text{min}}.
\end{align}
We take $\{v, m_h, m_S, \lambda_{HS}, \lambda_S\}$ as the input parameter set
in favor of the original one, $\{\mu_H^2, \mu_S^2, \lambda_H, \lambda_{HS}, \lambda_S\}$.
At the tree level, one gets
\begin{align}
\mu_H^2 = \frac{m_h^2}{2},\quad \mu_S^2 = -m_S^2+\frac{\lambda_{HS}}{2}v^2, 
\quad \lambda_H = \frac{m_h^2}{2v^2}.
\end{align}
In our numerical analyses, we take $\lambda_S=\lambda_S^{\text{min}}+0.1$ as adopted in Ref.~\cite{Curtin:2014jma}.

The tadpole conditions and scalar masses at one-loop level are calculated using~\cite{Jackiw:1974cv,VCW}
\begin{align}
V_{\rm CW}(\bar{m}_i^2) &= \sum_i n_i\frac{\bar{m}_i^4}{4(16\pi^2)}\left(\ln\frac{\bar{m}_i^2}{\bar{\mu}^2}-c_i\right),
\end{align}
which is regularized in the $\overline{\rm MS}$ scheme, where 
$\bar{m}_i$ are the background-field-dependent masses of the Higgs bosons ($H_{1,2}$),
the NG bosons ($G^0, G^\pm$), the weak gauge bosons ($W, Z$) and the top quark ($t$) with
$n_{H_1}=n_{H_2}=n_{G^0} = 1$, $n_{G^\pm} = 2$, $n_W=6$, $n_Z = 3$, $n_t=-12$, 
$c=3/2$ for the scalars and top quark while $c=5/6$ for the gauge bosons, and
$\bar{\mu}$ denotes the renormalization scale.
Note that $H_{1,2}$ are the admixtures of $h$ and $S$ occurring for field configurations 
other than the vacuum. 

We first describe the OS-like scheme in which the tree-level relations are not altered by the loop corrections~\cite{Kirzhnits:1976ts}. \footnote{For the genuine OS scheme in the SM with the singlet scalar, see, {\it e.g.}, Ref.~\cite{OSrSM}}.
To this end, the (finite) renormalization conditions are imposed as 
\begin{align}
\left\langle \frac{\partial (V_{\text{CW}}+V_{\text{CT}})}{\partial \varphi}\right\rangle = 0, \quad
\left\langle\frac{\partial^2 (V_{\text{CW}}+V_{\text{CT}})}{\partial \varphi^2}\right\rangle = 0,\quad
\left\langle\frac{\partial^2 (V_{\text{CW}}+V_{\text{CT}})}{\partial \varphi_S^2}\right\rangle = 0,
\label{OS}
\end{align}
where
\begin{align}
V_{\text{CT}} = -\frac{\delta \mu_H^2}{2}\varphi^2-\frac{\delta \mu_S^2}{2}\varphi_S^2.
\end{align} 
Note that the conditions (\ref{OS}) also fix $\bar{\mu}$ in addition to $\delta\mu_H^2$ and $\delta \mu_S^2$.
As a result, the renormalized one-loop effective potential takes the form
\begin{align}
V_1^{\text{(OS)}}(\varphi,\varphi_S)=\sum_in_i\frac{1}{4(16\pi^2)}
\left[
	\bar{m}_i^4\left(\ln\frac{\bar{m}_i^2}{m_i^2}-\frac{3}{2}\right)
	+2\bar{m}_i^2m_i^2
\right],\label{V1OS}
\end{align}
where $m_i^2 = \langle \bar{m}_i^2\rangle$. 
In this scheme, the NG bosons cause the IR divergence in the second condition in Eq.~(\ref{OS}). 
To circumvent it, their contributions should be treated with a special care. 
In this work, we adopt a prescription proposed in Refs.~\cite{Elias-Miro:2014pca,Martin:2014bca}. 
\footnote{The IR divergence issue can also be cured by using the on-shell Higgs mass rather than the zero-momentum defined Higgs mass~\cite{IRsolution2}.}
In this case, the resummed NG contributions take the form
\begin{align}
V_{\text{CW}}^{(G)}(\varphi) & = \frac{\bar{M}_{G^0}^4}{4(16\pi^2)}
\left(\ln\frac{\bar{M}_{G^0}^2}{\bar{\mu}^2}-\frac{3}{2} \right)
+2\cdot\frac{\bar{M}_{G^\pm}^4}{4(16\pi^2)}
\left(\ln\frac{\bar{M}_{G^\pm}^2}{\bar{\mu}^2}-\frac{3}{2} \right),
\label{Vcw_resumG}
\end{align}
where $\bar{M}_{G^{0,\pm}}^2 = \bar{m}_{G^{0,\pm}}^2+\bar{\Sigma}_G$ 
with $\bar{\Sigma}_G$ being the one-loop self-energy
of the NG bosons with vanishing external momenta,
\begin{align}
\bar{\Sigma}_G &= \frac{1}{16\pi^2}
\bigg[
	3\lambda_H\bar{m}_{H_1}^2\left(\ln\frac{\bar{m}_{H_1}^2}{\bar{\mu}^2}-1\right)
	+\frac{1}{2}\lambda_{HS}\bar{m}_{H_2}^2\left(\ln\frac{\bar{m}_{H_2}^2}{\bar{\mu}^2}-1\right)\nonumber\\
&\hspace{1.6cm}
	+\frac{3g_2^2}{2}\bar{m}_W^2\left(\ln\frac{\bar{m}_W^2}{\bar{\mu}^2}-\frac{1}{3}\right)
	+\frac{3(g_2^2+g_1^2)}{4}\bar{m}_Z^2\left(\ln\frac{\bar{m}_Z^2}{\bar{\mu}^2}-\frac{1}{3}\right)
	\nonumber \\
&\hspace{1.6cm}
	-6y_t^2\bar{m}_t^2\left(\ln\frac{\bar{m}_t^2}{\bar{\mu}^2}-1\right)
\bigg],\label{SigmaG}
\end{align}
where $g_{2,1}$ denote the SU(2)$_L$ and U(1)$_Y$ gauge couplings, respectively, and $y_t$ the top Yukawa coupling. The leading contribution comes from the top quark loop.
With this resummation prescription, the second derivative of Eq.~(\ref{Vcw_resumG}) evaluated in the vacuum is made finite, $m_{G^{0,\pm}}=0$.

Now we move on to discuss the one-loop corrected tadpole conditions and Higgs masses in the 
$\overline{\text{MS}}$ scheme.
In this case, we impose 
\begin{align}
T_h&=\left\langle\frac{\partial (V_0+V_{\text{CW}}) }{\partial \varphi}\right\rangle
=(-\mu_H^2+\lambda_H v^2)v+\left\langle\frac{\partial V_{\rm CW}}{\partial \varphi}\right\rangle=0,
\label{tad_MSbar}\\
m_h^2&=\left\langle\frac{\partial^2 (V_0+V_{\text{CW}})}{\partial \varphi^2}\right\rangle
=2\lambda_H v^2+\left\langle\frac{\partial^2 V_{\rm CW}}{\partial \varphi^2}\right\rangle
	-\frac{1}{v}\left\langle\frac{\partial V_{\rm CW}}{\partial \varphi}\right\rangle, \label{mh_MSbar} \\
m_S^2&=\left\langle\frac{\partial^2 (V_0+V_{\text{CW}})}{\partial \varphi_S^2}\right\rangle
=-\mu_S^2+\frac{\lambda_{HS}}{2}v^2+\left\langle\frac{\partial^2 V_{\rm CW}}{\partial \varphi_S^2}\right\rangle.
\label{mS_MSbar}
\end{align}
In Eq.~(\ref{mh_MSbar}), $\mu_H^2$ is eliminated by use of Eq.~(\ref{tad_MSbar}).
In contrast to the OS-like scheme, $m_h$ does not suffer from the IR divergence since 
$m_{G^{0,\pm}}\neq 0$ in the vacuum. 
We determine the parameters $\{\mu_H^2, \mu_S^2, \lambda_H\}$ by solving the above three conditions numerically. In our numerical analyses, $\bar{\mu}$ is varied from $m_t/2$ to $2m_t$
with $m_t=173.2$~GeV in order to quantify the scale uncertainty.

\section{Electroweak phase transition}\label{sec:ewpt}

For the EWBG scenario to work, the baryon-changing processes have to be sufficiently suppressed 
inside the expanding bubbles. The criterion for it is given by 
\begin{align}
\frac{v_C}{T_C}>\zeta_{\text{sph}}(T_C),
\end{align}
where $\zeta_{\text{sph}}(T_C)$ depends on the sphaleron configuration~\cite{sph}, 
the fluctuation determinants about it,
and so on~\cite{Ahriche:2007jp,Funakubo:2009eg,Patel:2011th,Fuyuto:2014yia,Ahriche:2014jna}.
In the current model, it is found that $\zeta_{\text{sph}}\simeq 1.1 - 1.2$~\cite{Fuyuto:2014yia},
where the one-loop effective potential with thermal resummation is used to evaluate the sphaleron energy.
It is thus $\xi$-dependent and must be revised in a gauge-invariant manner. We defer it to a future study.

To investigate the EWPT, we use the finite-$T$ one-loop effective potential given by~\cite{Dolan:1973qd}
\begin{align}
V_1^{T}(\varphi, \varphi_S; T) = \sum_{i}n_i
	\frac{T^4}{2\pi^2}I_{B,F}\left(\frac{\bar{m}_i^2}{T^2}\right),\quad
I_{B,F}(a^2) = \int_0^\infty dx~x^2\ln\Big(1\mp e^{-\sqrt{x^2+a^2}}\Big).
\label{V1T}
\end{align}
Since the perturbative expansion would break down at high temperatures, 
the dominant thermal pieces must be resummed. 
In this work, we adopt a prescription such that $\bar{m}_i^2$ appearing in the thermal function of 
$I_B(\bar{m}_i^2/T^2)$ 
are replaced with $\bar{m}_i^2+\Sigma_i(T)$ with $\Sigma_i(T)$ being the thermal masses (for a refined resummation method, see, {\it e.g.}, Ref.~\cite{Curtin:2016urg}). 
The explicit expressions of $\Sigma_i(T)$ can be found in Refs.~\cite{Espinosa:2011ax,Carrington:1991hz}

As pointed out in Ref.~\cite{Funakubo:2005pu}, two-step phase transitions have expanded
EWBG possibilities in models with singlet scalar extensions. 
In our case, the primary phase transition occurs from $(\varphi,\varphi_S)=(0, 0)$ to 
$(\varphi,\varphi_S)=(0, v_S^{\text{sym}})$, followed by the secondary transition to 
$(\varphi,\varphi_S)=(v, v_S^{\text{br}})$. 
The critical temperature, $T_C$, of the EWPT in standard gauge-dependent EWPT calculations 
is defined by the degenerate minima condition
\begin{align}
V_{\text{eff}}(0, v_{SC}^{\text{sym}}; T_C)=V_{\text{eff}}(v_C, v_{SC}^{\text{br}}; T_C),\label{Tc_STD}
\end{align}
where $v_C = \text{lim}_{T\uparrow T_C}v(T)$, 
$v_{SC}^{\text{br}} = \text{lim}_{T\uparrow T_C}v_S(T)$,
$v_{SC}^{\text{sym}} = \text{lim}_{T\downarrow T_C}v_S(T)$
with the uparrow (downarrow) being the limit taken from below (above) $T_C$.
We will determine $T_C$ and the VEVs using the effective potential at $T=0$ 
with the renormalization conditions explained above and Eq.~(\ref{V1T}) with the thermal resummation.

In the PRM scheme~\cite{Patel:2011th}, on the other hand, $T_C$ is determined so as to satisfy the NFK identity
expressed by
\begin{align}
\frac{\partial V_{\text{eff}}(\varphi)}{\partial \xi} = -C(\varphi, \xi)\frac{\partial V_{\text{eff}}(\varphi)}{\partial \varphi},
\end{align}
where $C(\varphi, \xi)$ is some functional. In the perturbative analysis, $V_{\text{eff}}$ and $C(\varphi, \xi)$ should be expanded in powers of $\hbar$:
\begin{align}
\begin{split}
V_{\text{eff}}(\varphi) &= V_0(\varphi)+\hbar V_1(\varphi) + \hbar^2 V_2(\varphi)+\cdots,\\
C(\varphi, \xi) & = c_0+\hbar c_1(\varphi, \xi)+\hbar^2 c_2(\varphi, \xi)+\cdots.
\end{split}
\end{align}
Since $c_0=0$ due to the $\xi$ independence of $V_0$,  
the identity to $\mathcal{O}(\hbar)$ is cast into the form
\begin{align}
\frac{\partial V_1}{\partial \xi} = -c_1\frac{\partial V_0}{\partial \varphi}.
\end{align}
Therefore, the $\xi$ dependence of $V_1$ disappears at the stationary points of $V_0$ rather than 
those of $V_1$. 

In the aforementioned two-step phase transition case, 
$T_C$ to $\mathcal{O}(\hbar)$ in the PRM is determined by 
\begin{align}
&
V_0(0, v_{S,\text{tree}}^{\text{sym}})+V_{\text{CW}}(0, v_{S,\text{tree}}^{\text{sym}})
+V_1^T(0, v_{S,\text{tree}}^{\text{sym}}; T_C)
\nonumber \\
&
=V_0(v_\text{tree}, 0)+V_{\text{CW}}(v_\text{tree}, 0)+V_1^T(v_\text{tree}, 0; T_C),\label{Tc_PRM}
\end{align}
where $v_{\text{tree}}=246$~GeV and
$v_{S,\text{tree}}^{\text{sym}}$ is the minimum of $V_0(0,\varphi_S)$.
Unlike the standard gauge-dependent calculations, the field values 
are fixed by the tree-level stationary points.
As a result, $T_C$ in this scheme becomes lower than those in the gauge-dependent calculations,
determined by Eq.~(\ref{Tc_STD}).
It is shown in Ref.~\cite{Chiang:2017nmu} that the $\bar{\mu}$ dependence in $V_{\text{CW}}$ can affect $T_C$ significantly.
This is due to the fact that 
in the ordinary gauge-dependent methods at one-loop level, the one-loop tadpole conditions, which are $\bar{\mu}$ dependent, are used in determining $T_C$.
As a result, the $\bar{\mu}$ dependences of $V_{\text{CW}}$ are partially cancelled.
In the PRM method, on the other hand, the tree-level tadpole conditions are used even at the one-loop order in order to satisfy the NFK identity, yielding the larger $\bar{\mu}$ dependences.
In Ref.~\cite{Chiang:2017nmu}, using renormalization group equations, 
$(V_0+V_{\text{CW}})$ is made $\bar{\mu}$-independent up to higher-order corrections. 
However, we still have degrees of freedom to choose an input scale for the running parameters
to which $T_C$ is vulnerable.
The fundamental solution for it may require higher order corrections that are missing here.
In the current analysis, we do not elaborate a more refined calculation and just
vary $\bar{\mu}$ from $m_t/2$ to $2m_t$ in order to estimate the scale uncertainty of $T_C$
as in the $\overline{\text{MS}}$ scheme.

In the PRM scheme, the VEVs at $T_C$ are determined by the minima of the HT potential
defined by
\begin{align}
V^{\text{HT}}(\varphi, \varphi_S) = V_0(\varphi, \varphi_S)
+\frac{1}{2}\Sigma_H(T)\varphi^2+\frac{1}{2}\Sigma_S(T)\varphi_S^2,
\end{align}
where $\Sigma_H(T)$ and $\Sigma_S(T)$ are the thermal masses of $H$ and $S$, respectively~\cite{Espinosa:2011ax}.
The HT potential is manifestly $\xi$ independent, thanks to the $\xi$ independence of the thermal masses as mentioned in Introduction. 
Because of this nice property, it is possible to obtain the gauge-invariant $T_C$ and VEVs
by solely using the potential.
Application of the HT scheme to the singlet-extended SMs can be found in Ref.~\cite{HT_xSM}.

\section{Results}\label{sec:results}

Here we conduct the numerical analysis. 
The free parameters in this model are $m_S$, $\lambda_{HS}$ and $\lambda_S$.
In the current study, we take $m_S= m_h/2$ that is consistent with the DM phenomenology
 \footnote{It is well known that the DM relic density is lower than the 
observed value in parameter space consistent with the strong first-order EWPT.
For a recent study of DM in this model, see, {\it e.g.}, Ref.~\cite{DM_rSM_recent}.}
and $\lambda_S = \lambda_S^{\text{min}}+0.1$, and thus $\lambda_{HS}$ is the only parameter we vary. 
We focus mostly on the parameter space where $v_C/T_C\simeq1$ realized by the two-step EWPT 
associate with the tree-level potential barrier. 
In this case, the range of $\lambda_{HS}$ is also more or less fixed.

\begin{figure}[t]
\center
\includegraphics[width=8cm]{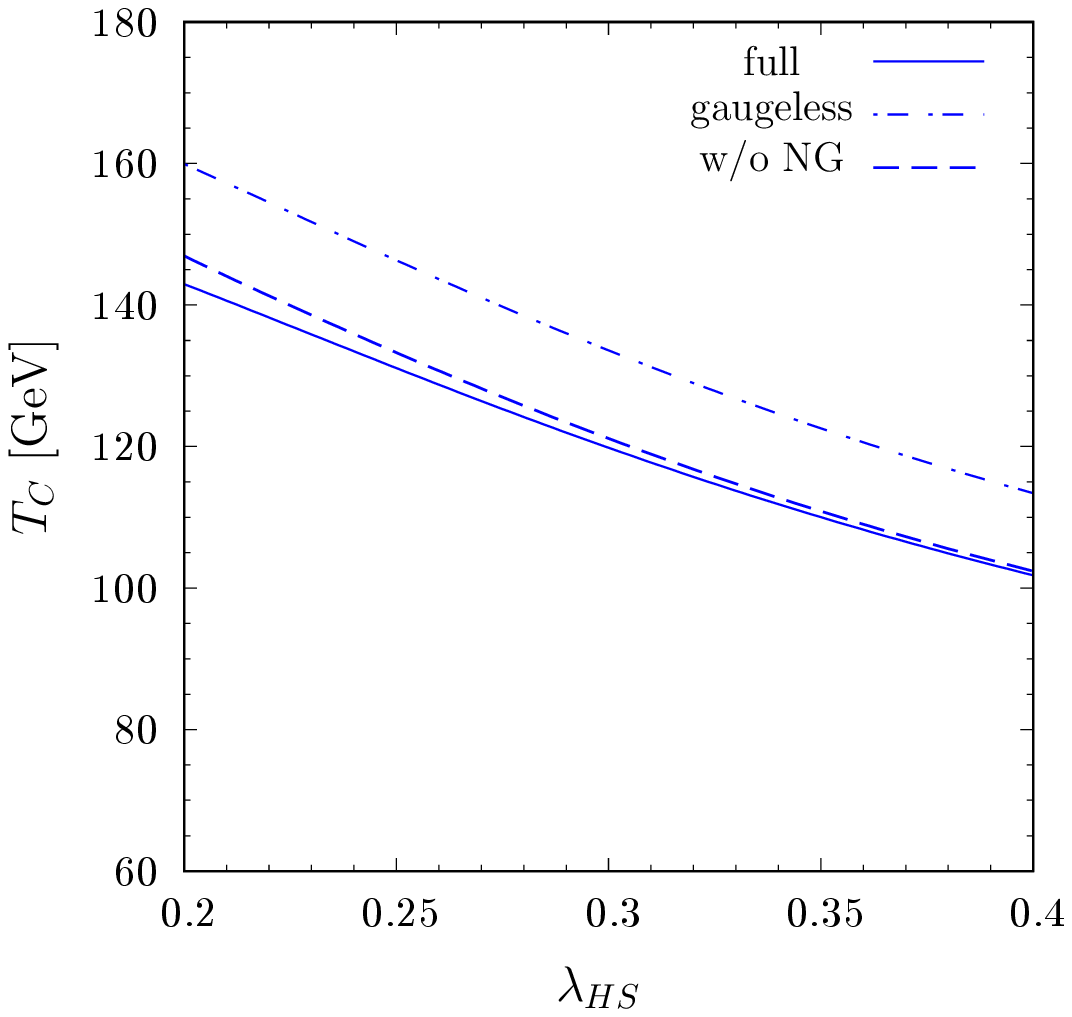}
\hspace{0.2cm}
\includegraphics[width=8cm]{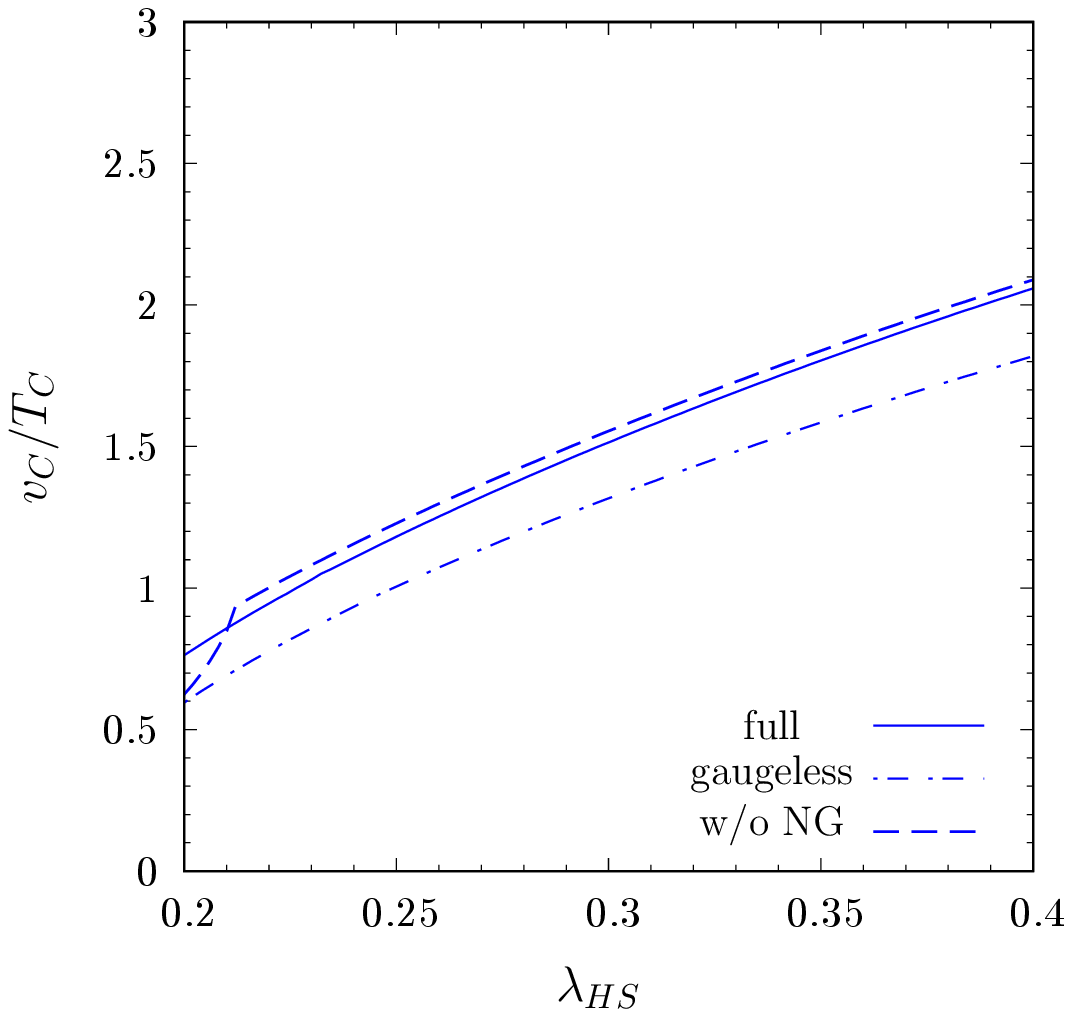}
\caption{Impacts of the thermal gauge bosons (dot-dashed curves) and NG bosons (dashed curves) on $T_C$ (left) and $v_C/T_C$ (right) as a function of $\lambda_{HS}$. The solid curves include 
both contributions. Here the OS-like scheme is used.}
\label{fig:STD}
\end{figure}

In Fig.~\ref{fig:STD}, we study the EWPT in two approximations.
One is the calculation without including the thermal gauge boson loops, denoted as ``gaugeless" and depicted by the dot-dashed curves,
\footnote{We have confirmed that the zero temperature gauge boson loops have little effect on EWPT.}
and the other is the one without the NG boson contributions, denoted as ``w/o NG" and depicted by the dashed curves.  The solid curves labeled by ``full" includes both of them.  Here the OS-like renormalization scheme is adopted.  The left and right plots show $T_C$ and $v_C/T_C$ as functions of $\lambda_{HS}$, respectively.
One can see that the thermal gauge boson loops have a $(12-17)\%$ effect on $T_C$
and $(12-22)\%$ on $v_C/T_C$. 
What is remarkable here is that the importance of the gauge boson loops persists even if 
the tree-potential barrier exists.
As mentioned in Introduction, the figures are not necessarily equivalent to the $\xi$ dependence itself,
but it is expected that the larger percentages naively correspond to a greater possibility of 
the $\xi$ artifact. 
Formally, the $\xi$ dependence comes from the next order in the perturbative expansion so that 
its magnitude is not so large as long as $\xi$ is assumed to be an $\mathcal{O}(1)$ parameter, which may not be justified {\it a priori} though. 
As discussed in Ref.~\cite{Chiang:2017zbz}, however, even if the $\xi$ dependence on $T_C$ 
is a few \%, it cannot guarantee that the bubble nucleation temperature or gravitational waves generated during the first-order phase transitions also have similar $\xi$ dependences.
Actually, the gravitational wave spectrum in a U(1)$_{B-L}$ model discussed in Ref.~\cite{Chiang:2017zbz} 
can change by one order magnitude when varying $\xi$ from 0 to 5.
Having this in mind, the results shown in Fig.~\ref{fig:STD} motivate us to conduct further investigations 
in the current model as well.
The quantification of the $\xi$ dependence on the EWPT using the general $R_\xi$ gauge will be given elsewhere.

We also find that the NG boson effects are $(0.6-2.7)\%$ in $T_C$ and $(1.5-18)\%$ 
in $v_C/T_C$, respectively.  
Note that the effect becomes more pronounced if the thermal potential barrier dominates over the 
tree-level potential barrier, which occurs when $\lambda_{HS}\simeq 0.21$ and below, as shown
by the bend in the dashed curve of the right panel. Otherwise, the effect is typically at a few \% level.

\begin{figure}[t]
\center
\includegraphics[width=8cm]{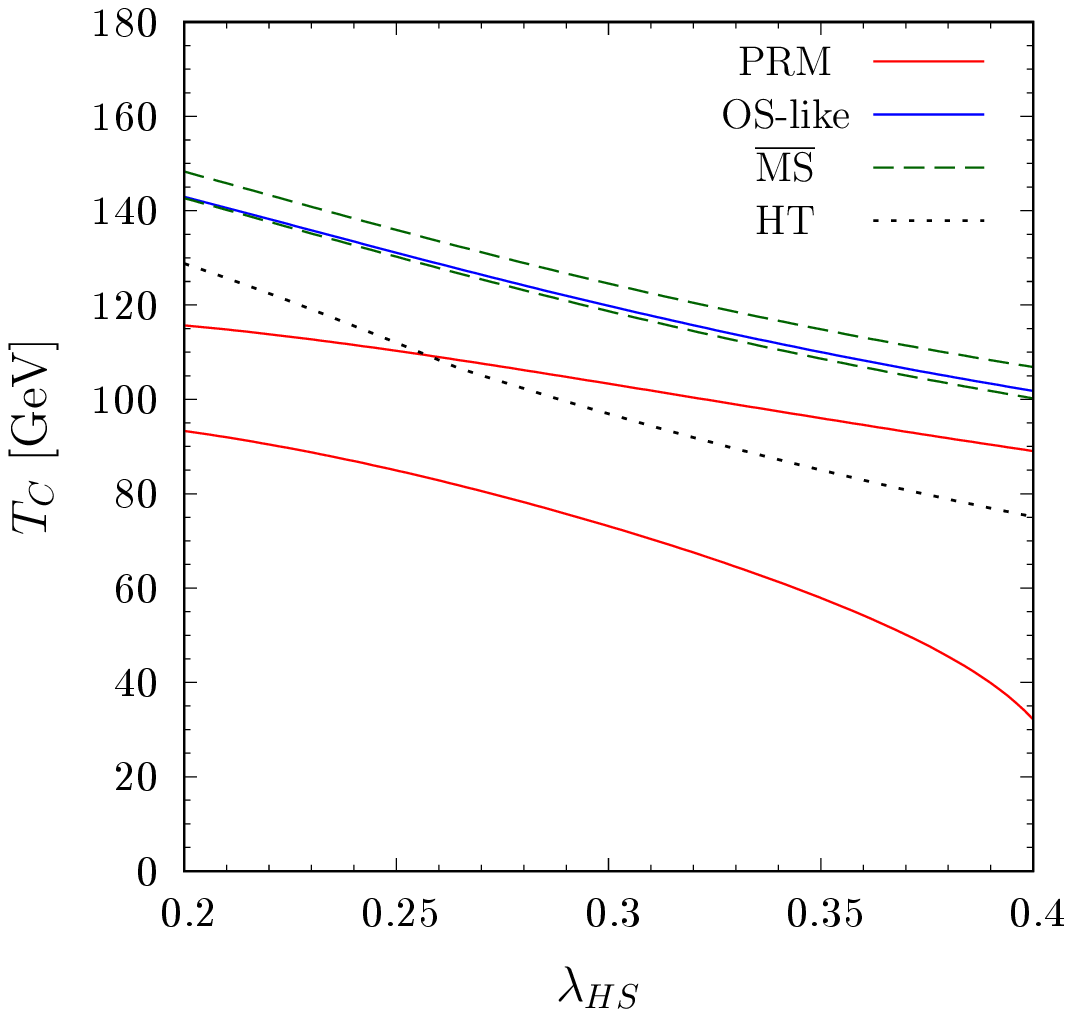}
\hspace{0.2cm}
\includegraphics[width=8cm]{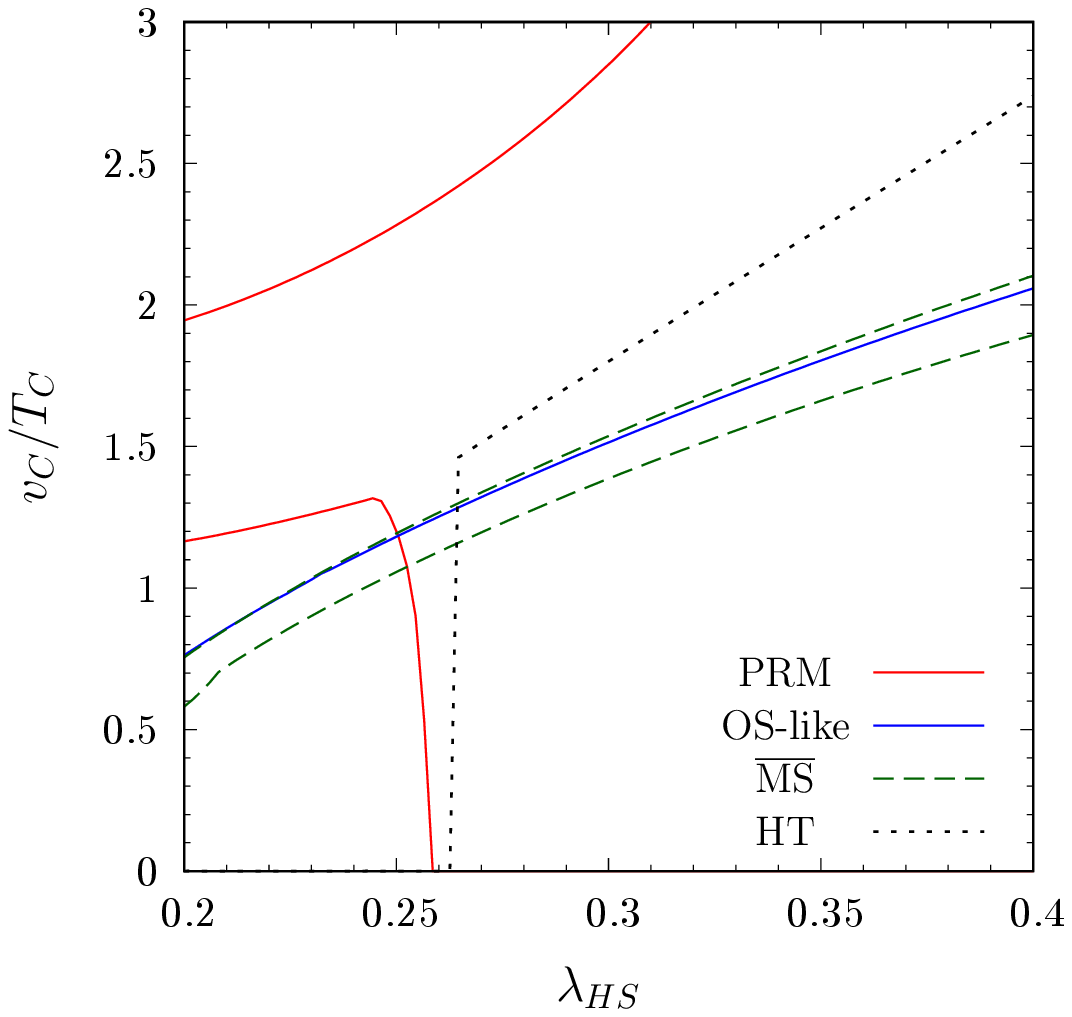}
\caption{Comparisons among the various calculation methods: PRM (red-solid), OS-like scheme with 
the NG resummation (blue-solid),
$\overline{\text{MS}}$ (green-dash), and HT (black-dot), respectively. For PRM and $\overline{\text{MS}}$, $\bar{\mu}$ is varied from $m_t/2$ to $2m_t$.}
\label{fig:all}
\end{figure}

Now we investigate the scheme dependence using the other methods:
PRM, $\overline{\text{MS}}$, and HT schemes. 
The numerical results are summarized in Fig.~\ref{fig:all}. 
The colors and styles of the curves are as follows: PRM scheme (red-solid), OS-like scheme with the NG resummation (blue-solid), $\overline{\text{MS}}$ scheme (green-dash), and HT scheme (black-dot). 
For the PRM and $\overline{\text{MS}}$ schemes, $\bar{\mu}$ is varied from $m_t/2$ and $2m_t$.
We find the following:

\begin{itemize}
\item[1.] 
The OS-like and $\overline{\text{MS}}$ schemes show a nice agreement between each other within the
scale uncertainties in the $\overline{\text{MS}}$ scheme which are $(3.8-6.2)\%$ in $T_C$ and $(10-23)\%$ in $v_C/T_C$, respectively.
Here the upper (lower) curve in $T_C$ corresponds to the case with $\bar{\mu}=2m_t~(m_t/2)$, and the other way around for $v_C/T_C$.
One can find that the two results get closer if $\bar{\mu}=m_t/2$ is taken.
For the commonly used choice in the literature, $\bar{\mu}=m_t$, on the other hand, 
$T_C$ ($v_C/T_C$) in the $\overline{\text{MS}}$ scheme is larger (smaller) than that 
in the OS-like scheme by $\sim 1.5$\% $(2.7-9.5)\%$.
In any case, the relatively large scale uncertainties, especially in $v_C/T_C$ in the $\overline{\text{MS}}$ scheme indicates the necessity of higher-order corrections.

\item[2.] The PRM scheme gives qualitatively the same behavior of $T_C$ against $\lambda_{HS}$
as in the OS-like and $\overline{\text{MS}}$ schemes; namely, $T_C$ gets smaller as $\lambda_{HS}$ increases. 
Here the upper (lower) curve in $T_C$ is for $\bar{\mu}=m_t/2~(2m_t)$, and the other way around for $v_C/T_C$.
One can see that this scheme is subject to more scale uncertainties as mentioned in Sec.~\ref{sec:ewpt}.
In spite of this, one of the universal features of this scheme is that $T_C$ is lower than the gauge-dependent $T_C$, which is the consequence of the different determination of $T_C$; {\it i.e.}, the degenerate point is away from the minimum of the one-loop effective potential, 
and hence the degeneracy 
occurs at a lower $T$.

Because of the lower $T_C$, $v_C/T_C$ is enhanced compared to those of the OS-like and $\overline{\text{MS}}$ schemes except around $\bar{\mu}=m_t/2$, where
$v_C$ becomes zero for $\lambda_{HS}\gtrsim 0.25$ since $T_C$ in PRM 
gets larger than that in HT.  
Nevertheless, we conclude that there is no significant inconsistency
between the PRM and other schemes within the theoretical uncertainties.
In any case, a more refined calculation with higher-order corrections such as at 
$\mathcal{O}(\hbar^2)$ and the daisy diagrams
is indispensable for a quantitative analysis.

\item[3.] The critical temperature $T_C$ in the HT scheme can be smaller than those in the OS-like and $\overline{\text{MS}}$ schemes by about $(10-30)$~GeV. 
We also find that the EWPT in the HT scheme is not first-order for $\lambda_{HS}\lesssim 0.26$.
Moreover, even if it becomes first order, $v_C/T_C$ is overestimated compared to the other two schemes, which signifies the importance of the one-loop corrections. 
\end{itemize}

Before closing this section, we briefly comment on the Landau pole issue in this scenario.
In most EWBG scenarios, the region of $v_C/T_C>1$ is not compatible with the absence of the
Landau pole up to the Planck scale ($\sim 10^{19}$~GeV). 
In the above scenario, in contrast, it is found that all the couplings in the model are less than $4\pi$ all the way to the Planck scale for $\lambda_{HS}\lesssim 0.33$ using the one-loop renormalization group equations.

\section{Conclusion and discussions}\label{sec:conclusion}

We have revisited EWPT in the singlet-extended SM using several calculation methods
to study the scheme dependence. 
In the OS-like scheme, the NG bosons must be taken with a special care in order to avoid the IR divergence.
Here we adopted the NG resummation method recently proposed in Refs.~\cite{Martin:2014bca,Elias-Miro:2014pca} and quantified such a resummation effect on first-order EWPT.
It is found that the effect can get pronounced if the potential barrier is governed mainly
by the thermal cubic loops rather than the tree-potential structure. If not, the effect is typically at $\sim 1$\% level. In addition, we numerically studied the impacts of the thermal gauge boson loops on $v_C/T_C$ and found that such loops had a $(12-22)\%$ effect on $v_C/T_C$ even when the 
tree-level potential barrier existed.
This motivates us to conduct the precise quantification of the $\xi$ dependence 
using the general $R_\xi$ gauge in a future study.

We also found that the results in the OS-like and $\overline{\text{MS}}$ schemes
showed a nice agreement within the scale uncertainties which are $(3.8-6.2)\%$ in $T_C$ and $(10-23)\%$ in $v_C/T_C$. 
Our numerical studies also clarified that 
$T_C$ and the corresponding VEVs against $\lambda_{HS}$ in the gauge-invariant PRM method 
were qualitatively consistent with those in the above gauge-dependent schemes within the 
rather large theoretical uncertainties.
Regardless of the gauge-dependent or -independent methods, we found that 
the scale uncertainties in $v_C/T_C$ were more than about 10\%, 
suggesting that higher-order corrections could be potentially important.

\begin{acknowledgments}
This work was supported in part by the Ministry of Science and Technology of Taiwan under Grant Nos. 104-2628-M-008-004-MY4 and 104-2811-M-008-056, and IBS under the project code, IBS-R018-D1.
\end{acknowledgments}



\begin{thebibliography}{99}


\bibitem{Patrignani:2016xqp} 
  C.~Patrignani {\it et al.} [Particle Data Group],
  Chin.\ Phys.\ C {\bf 40}, no. 10, 100001 (2016).

\bibitem{Sakharov:1967dj} 
  A.~D.~Sakharov,
  Pisma Zh.\ Eksp.\ Teor.\ Fiz.\  {\bf 5}, 32 (1967)
  [JETP Lett.\  {\bf 5}, 24 (1967)]
  [Sov.\ Phys.\ Usp.\  {\bf 34}, no. 5, 392 (1991)]
  [Usp.\ Fiz.\ Nauk {\bf 161}, no. 5, 61 (1991)].
  
\bibitem{125h}  
  G.~Aad {\it et al.} [ATLAS Collaboration],
  Phys.\ Lett.\ B {\bf 716}, 1 (2012);~
%
  S.~Chatrchyan {\it et al.} [CMS Collaboration],
  Phys.\ Lett.\ B {\bf 716}, 30 (2012).

\bibitem{ewbg}
  V.~A.~Kuzmin, V.~A.~Rubakov and M.~E.~Shaposhnikov,
  Phys.\ Lett.\ B {\bf 155} (1985) 36.
For reviews on electroweak baryogenesis, see
A.~G.~Cohen, D.~B.~Kaplan and A.~E.~Nelson,
Ann.\ Rev.\ Nucl.\ Part.\ Sci.\  {\bf 43} (1993) 27;~
%
M.~Quiros,
Helv.\ Phys.\ Acta {\bf 67} (1994) 451;~
%
V.~A.~Rubakov and M.~E.~Shaposhnikov,
Usp.\ Fiz.\ Nauk {\bf 166} (1996) 493;~
%
K.~Funakubo,
Prog.\ Theor.\ Phys.\  {\bf 96} (1996) 475;~
%
M.~Trodden,
Rev.\ Mod.\ Phys.\  {\bf 71} (1999) 1463;~
%
W.~Bernreuther,
Lect.\ Notes Phys.\  {\bf 591} (2002) 237;~
%
  J.~M.~Cline,
  [arXiv:hep-ph/0609145];~
  D.~E.~Morrissey and M.~J.~Ramsey-Musolf,
  New J.\ Phys.\  {\bf 14}, 125003 (2012);~
%
%
  T.~Konstandin,
  Phys.\ Usp.\  {\bf 56} (2013) 747
   [Usp.\ Fiz.\ Nauk {\bf 183} (2013) 785].

\bibitem{lattice}
  K.~Kajantie, M.~Laine, K.~Rummukainen and M.~E.~Shaposhnikov,
  Phys.\ Rev.\ Lett.\  {\bf 77}, 2887 (1996);~
  K.~Rummukainen, M.~Tsypin, K.~Kajantie, M.~Laine and M.~E.~Shaposhnikov,
  Nucl.\ Phys.\ B {\bf 532}, 283 (1998);~
  F.~Csikor, Z.~Fodor and J.~Heitger,
  Phys.\ Rev.\ Lett.\  {\bf 82}, 21 (1999);~
  Y.~Aoki, F.~Csikor, Z.~Fodor and A.~Ukawa,
  Phys.\ Rev.\ D {\bf 60}, 013001 (1999).
            
\bibitem{EWPT_rSM}      
  J.~R.~Espinosa and M.~Quiros,
  Phys.\ Rev.\ D {\bf 76}, 076004 (2007);~
%
  S.~Profumo, M.~J.~Ramsey-Musolf and G.~Shaughnessy,
  JHEP {\bf 0708}, 010 (2007);~
%
  D.~J.~H.~Chung, A.~J.~Long and L.~T.~Wang,
  Phys.\ Rev.\ D {\bf 87}, no. 2, 023509 (2013);~
%
  N.~Craig, H.~K.~Lou, M.~McCullough and A.~Thalapillil,
  JHEP {\bf 1602}, 127 (2016);~
%
  S.~Ghosh, A.~Kundu and S.~Ray,
  Phys.\ Rev.\ D {\bf 93}, no. 11, 115034 (2016);~
%
  T.~Tenkanen, K.~Tuominen and V.~Vaskonen,
  JCAP {\bf 1609}, no. 09, 037 (2016);~
%
  P.~H.~Ghorbani,
  JHEP {\bf 1708}, 058 (2017);~
%
  L.~Marzola, A.~Racioppi and V.~Vaskonen,
  Eur.\ Phys.\ J.\ C {\bf 77}, no. 7, 484 (2017);~
%
  G.~Kurup and M.~Perelstein,
  Phys.\ Rev.\ D {\bf 96}, no. 1, 015036 (2017);~
%
  B.~Jain, S.~J.~Lee and M.~Son,
  arXiv:1709.03232 [hep-ph];~
%
  K.~Ghorbani and P.~H.~Ghorbani,
  arXiv:1804.05798 [hep-ph].

\bibitem{Espinosa:2011ax} 
  J.~R.~Espinosa, T.~Konstandin and F.~Riva,
  Nucl.\ Phys.\ B {\bf 854}, 592 (2012).
%

\bibitem{Curtin:2014jma} 
  D.~Curtin, P.~Meade and C.~T.~Yu,
  JHEP {\bf 1411}, 127 (2014).

\bibitem{Curtin:2016urg} 
  D.~Curtin, P.~Meade and H.~Ramani,
  arXiv:1612.00466 [hep-ph].

\bibitem{HT_xSM}
  V.~Vaskonen,
  Phys.\ Rev.\ D {\bf 95}, no. 12, 123515 (2017).

\bibitem{DM_rSM}   
  V.~Silveira and A.~Zee,
  Phys.\ Lett.\  {\bf 161B}, 136 (1985);~
%
  J.~McDonald,
  Phys.\ Rev.\ D {\bf 50}, 3637 (1994);~
%
  C.~P.~Burgess, M.~Pospelov and T.~ter Veldhuis,
  Nucl.\ Phys.\ B {\bf 619}, 709 (2001);~
%
  D.~O'Connell, M.~J.~Ramsey-Musolf and M.~B.~Wise,
  Phys.\ Rev.\ D {\bf 75}, 037701 (2007);~
%
  V.~Barger, P.~Langacker, M.~McCaskey, M.~J.~Ramsey-Musolf and G.~Shaughnessy,
  Phys.\ Rev.\ D {\bf 77}, 035005 (2008);~
  %
  M.~Gonderinger, Y.~Li, H.~Patel and M.~J.~Ramsey-Musolf,
  JHEP {\bf 1001}, 053 (2010);~
%
  W.~L.~Guo and Y.~L.~Wu,
  JHEP {\bf 1010}, 083 (2010);~
%
  A.~Bandyopadhyay, S.~Chakraborty, A.~Ghosal and D.~Majumdar,
  JHEP {\bf 1011}, 065 (2010);~
%
  S.~Profumo, L.~Ubaldi and C.~Wainwright,
  Phys.\ Rev.\ D {\bf 82}, 123514 (2010);~
%
  Y.~Mambrini,
  Phys.\ Rev.\ D {\bf 84}, 115017 (2011);~
%
  A.~Biswas and D.~Majumdar,
  Pramana {\bf 80}, 539 (2013);~
%
  J.~M.~Cline, K.~Kainulainen, P.~Scott and C.~Weniger,
  Phys.\ Rev.\ D {\bf 88}, 055025 (2013)
  Erratum: [Phys.\ Rev.\ D {\bf 92}, no. 3, 039906 (2015)];~
%
  A.~Falkowski, C.~Gross and O.~Lebedev,
  JHEP {\bf 1505}, 057 (2015);~
%
  S.~Bhattacharya, P.~Poulose and P.~Ghosh,
  JCAP {\bf 1704}, no. 04, 043 (2017);~
%
  J.~A.~Casas, D.~G.~Cerdeño, J.~M.~Moreno and J.~Quilis,
  JHEP {\bf 1705}, 036 (2017).

\bibitem{DM_rSM_recent}
  P.~Athron {\it et al.} [GAMBIT Collaboration],
  Eur.\ Phys.\ J.\ C {\bf 77}, no. 8, 568 (2017);~
%
  P.~Athron, J.~M.~Cornell, F.~Kahlhoefer, J.~Mckay, P.~Scott and S.~Wild,
  arXiv:1806.11281 [hep-ph].

\bibitem{EWPT_DM_rSM}
  S.~Das, P.~J.~Fox, A.~Kumar and N.~Weiner,
  JHEP {\bf 1011}, 108 (2010);~
%
  D.~J.~H.~Chung and A.~J.~Long,5193
  Phys.\ Rev.\ D {\bf 84}, 103513 (2011);~
%
  J.~M.~Cline and K.~Kainulainen,
  JCAP {\bf 1301}, 012 (2013);~
%
  J.~M.~Cline, K.~Kainulainen, P.~Scott and C.~Weniger,
  Phys.\ Rev.\ D {\bf 88} (2013) 055025
   Erratum: [Phys.\ Rev.\ D {\bf 92} (2015) no.3,  039906];~
%
  T.~Alanne, K.~Tuominen and V.~Vaskonen,
  Nucl.\ Phys.\ B {\bf 889}, 692 (2014);~
%
  M.~Chala, G.~Nardini and I.~Sobolev,
  Phys.\ Rev.\ D {\bf 94}, no. 5, 055006 (2016).
  
\bibitem{Jackiw:1974cv} 
  R.~Jackiw,
  Phys.\ Rev.\ D {\bf 9}, 1686 (1974).
%
\bibitem{Dolan:1974gu} 
  L.~Dolan and R.~Jackiw,
  Phys.\ Rev.\ D {\bf 9}, 2904 (1974).

\bibitem{Patel:2011th} 
  H.~H.~Patel and M.~J.~Ramsey-Musolf,
  JHEP {\bf 1107}, 029 (2011).

\bibitem{Garny:2012cg} 
  M.~Garny and T.~Konstandin,
  JHEP {\bf 1207}, 189 (2012).

\bibitem{Wainwright:2012zn} 
  C.~L.~Wainwright, S.~Profumo and M.~J.~Ramsey-Musolf,
  Phys.\ Rev.\ D {\bf 86}, 083537 (2012).

\bibitem{Martin:2014bca} 
  S.~P.~Martin,
  Phys.\ Rev.\ D {\bf 90}, no. 1, 016013 (2014).

\bibitem{Elias-Miro:2014pca} 
  J.~Elias-Miro, J.~R.~Espinosa and T.~Konstandin,
  JHEP {\bf 1408}, 034 (2014).

\bibitem{IRsolution2}
  J.~M.~Cline and P.~A.~Lemieux,
  Phys.\ Rev.\ D {\bf 55}, 3873 (1997);~
%
  C.~Delaunay, C.~Grojean and J.~D.~Wells,
  JHEP {\bf 0804} (2008) 029;~
%
  J.~M.~Cline, K.~Kainulainen and M.~Trott,
  JHEP {\bf 1111}, 089 (2011).
  
  
\bibitem{Nielsen:1975fs} 
  N.~K.~Nielsen,
  Nucl.\ Phys.\ B {\bf 101}, 173 (1975).
  
\bibitem{Fukuda:1975di} 
  R.~Fukuda and T.~Kugo,
  Phys.\ Rev.\ D {\bf 13}, 3469 (1976).

\bibitem{Chiang:2017nmu} 
  C.~W.~Chiang, M.~J.~Ramsey-Musolf and E.~Senaha,
  Phys.\ Rev.\ D {\bf 97}, no. 1, 015005 (2018).


\bibitem{VCW}
  S.~R.~Coleman and E.~J.~Weinberg,
  Phys.\ Rev.\ D {\bf 7}, 1888 (1973);~
%
  S.~Weinberg,
  Phys.\ Rev.\ D {\bf 7}, 2887 (1973).
  
\bibitem{Kirzhnits:1976ts} 
  D.~A.~Kirzhnits and A.~D.~Linde,
  Annals Phys.\  {\bf 101}, 195 (1976).

\bibitem{OSrSM}
  S.~Kanemura, M.~Kikuchi and K.~Yagyu,
  Nucl.\ Phys.\ B {\bf 917}, 154 (2017);~
%
  S.~Kanemura, M.~Kikuchi, K.~Sakurai and K.~Yagyu,
  Phys.\ Rev.\ D {\bf 96}, no. 3, 035014 (2017);~
%
  G.~Ria and D.~Meloni,
  Eur.\ Phys.\ J.\ C {\bf 78}, no. 3, 270 (2018).

\bibitem{sph}
  N.~S.~Manton,
  Phys.\ Rev.\ D {\bf 28}, 2019 (1983);~
%
  F.~R.~Klinkhamer and N.~S.~Manton,
  Phys.\ Rev.\ D {\bf 30}, 2212 (1984).

\bibitem{Ahriche:2007jp} 
  A.~Ahriche,
  Phys.\ Rev.\ D {\bf 75}, 083522 (2007).

\bibitem{Funakubo:2009eg} 
  K.~Funakubo and E.~Senaha,
  Phys.\ Rev.\ D {\bf 79}, 115024 (2009).

\bibitem{Fuyuto:2014yia} 
  K.~Fuyuto and E.~Senaha,
  Phys.\ Rev.\ D {\bf 90}, no. 1, 015015 (2014).

\bibitem{Ahriche:2014jna} 
  A.~Ahriche, T.~A.~Chowdhury and S.~Nasri,
  JHEP {\bf 1411}, 096 (2014)

\bibitem{Dolan:1973qd} 
  L.~Dolan and R.~Jackiw,
  Phys.\ Rev.\ D {\bf 9}, 3320 (1974).

\bibitem{Carrington:1991hz} 
  M.~E.~Carrington,
  Phys.\ Rev.\ D {\bf 45}, 2933 (1992).

\bibitem{Funakubo:2005pu} 
  K.~Funakubo, S.~Tao and F.~Toyoda,
  Prog.\ Theor.\ Phys.\  {\bf 114}, 369 (2005).

  
\bibitem{Chiang:2017zbz} 
  C.~W.~Chiang and E.~Senaha,
  Phys.\ Lett.\ B {\bf 774}, 489 (2017).
  

\end{thebibliography}
\end{document}